\documentclass[twocolumn,tighten,twocolappendix]{aastex63}
\usepackage[T1]{fontenc}
\usepackage{graphicx}
\usepackage{amsmath}
\usepackage{amsfonts}
\usepackage{amssymb}
\usepackage{natbib}
\usepackage{textcomp}
\usepackage{gensymb}
\begin{document}

\title{From Formation to Disruption: Observing the Multiphase Evolution of a Solar Flare Current Sheet}

\correspondingauthor{L. P. Chitta}
\email{chitta@mps.mpg.de}

\author[0000-0002-9270-6785]{L. P. Chitta}
\affiliation{Max Planck Institute for Solar System Research, Justus-von-Liebig-Weg 3, D-37077 G\"ottingen, Germany}

\author[0000-0003-3621-6690]{E. R. Priest}
\affiliation{St Andrews University, Mathematics Institute, St Andrews KY16 9SS, UK}

\author[0000-0003-2837-7136]{X. Cheng}
\affiliation{School of Astronomy and Space Science, Nanjing University, Nanjing 210023, People's Republic of China}
\affiliation{Max Planck Institute for Solar System Research, Justus-von-Liebig-Weg 3, D-37077 G\"ottingen, Germany}

\begin{abstract}
A current sheet, where magnetic energy is liberated through reconnection and is converted to other forms, is thought to play the central role in solar flares, the most intense explosions in the heliosphere. However, the evolution of a current sheet and its subsequent role in flare-related phenomena such as particle acceleration is poorly understood. Here we report observations obtained with NASA's Solar Dynamics Observatory that reveal a multiphase evolution of a current sheet in the early stages of a solar flare, from its formation to quasi-stable evolution and disruption. Our observations have implications for the understanding of the onset and evolution of reconnection in the early stages of eruptive solar flares. 
\end{abstract}


\section{Introduction\label{sec:int}}

The magnetic energy necessary to power a solar flare is thought to be liberated through reconnection in the solar corona \citep[][]{1994Natur.371..495M,2002A&ARv..10..313P,2013NatPh...9..489S}. Observations of coronal magnetic fields during solar flares further support the idea of reconnection \citep[][]{2020Sci...367..278F}. A widely accepted scenario for the main phase of an eruptive two-ribbon flare consists of three types of structure. The lowest structure is a pair of bright chromospheric ribbons that are joined by an arcade of flare loops. During the rise phase one loop brightens up with bright knots at its feet and then they expand to form the ribbons and the flare loop arcade. During the main phase the ribbons move apart while the arcade of loops rise \citep[][]{1964NASSP..50..451C,1966Natur.211..695S,1974SoPh...34..323H,1976SoPh...50...85K}. Above this structure there lies a current sheet where much of the energy is released by reconnection, which starts at one location and so powers the initial loop and footpoint knots \citep[][]{1985RPPh...48..955P,2000JGR...105.2375L,2015SSRv..194..237L,2017SoPh..292...25P}. Later the reconnection spreads throughout the current sheet as the flare proceeds. 

Depending on whether the arcade is viewed from the side (face-on) or the ends (edge-on), the current sheet is observed from the side \citep[where supra-arcade downflows are seen during the main phase][]{1999ApJ...519L..93M,2003SoPh..217..247I,2012ApJ...747L..40S} or from the end as a thin structure protruding out into the corona \citep[e.g.][]{2002ApJ...575.1116C,2003JGRA..108.1440W,2008ApJ...686.1372C,2010ApJ...723L..28L,2010ApJ...722..329S,2018ApJ...853L..18Y,2018ApJ...854..122W,2018ApJ...866...64C,2018ApJ...868..148L}. Above the flare loops and current sheet, a flux rope erupts outward, starting slowly during the preflare phase and suddenly accelerating at the impulsive phase when reconnection is initiated \citep[e.g.][]{2013MNRAS.434.1309L,2013ApJ...767..168L,2020ApJ...894...85C}.

Post-impulsive phase current sheets are rather stable and persist for a long duration of up to several hours \citep[][]{2015SSRv..194..237L,2020ApJ...900..192F}. Macroscopic properties of these stable current sheets such as plasma flows and signatures of turbulence have been investigated using imaging and spectroscopic observations \citep[e.g.][]{2008ApJ...686.1372C,2018ApJ...854..122W,2018ApJ...866...64C,2020ApJ...900...17Y}. However, the nature of current sheets, their formation and subsequent evolution during the impulsive rise phase of a flare is limited \citep[][]{2013MNRAS.434.1309L,2018ApJ...853L..18Y,2019SciA....5.7004G}. 

Here we report observations of such a rare event, which provides a clearer picture of the evolution of a dynamic current sheet during the impulsive phase of a flare. Furthermore, the event is such that part of the current sheet is viewed edge-on and part from the side. In addition, the observations are extensive during the impulsive phase, and so enable us to obtain clues about how the current sheet develops and evolves.

\section{Observations\label{sec:obs}}
The Sun produced one of the largest solar flares in the last three years, and the most powerful one yet in the new solar cycle 25, on 2020 November 29 around UT\,13:00. Based on the soft X-ray flux detected by the 1-8\,\AA\ band of the Geostationary Operational Environmental Satellite (GOES), the events is classified as an M4.4-class flare. The Atmospheric Imaging Assembly \citep[AIA;][]{2012SoPh..275...17L} onboard the Solar Dynamics Observatory \citep[][]{2012SoPh..275....3P} observed this flare at the south-eastern limb. The flare is produced in active region AR 12790 when it was still behind the limb. For this reason, the extreme ultraviolet (EUV) filters on AIA could observe only the top of the flaring loop arcade. 

Here we focus on a 1-hour time sequence of AIA data from UT\,12:30 to UT\,13:30 covering the impulsive phase of the flare. In particular, we use AIA data recorded by six of its EUV filters. The central wavelengths of these filters, dominant ion species contributing to the emission in the respective passbands and the formation temperature of the ions are 94\,\AA\ (Fe\,{\sc x}: 6.05; Fe\,{\sc xviii}: 6.85), 131\,\AA\ (Fe\,{\sc viii}: 5.6; Fe\,{\sc xxi}: 7.05),  171\,\AA\ (Fe\,{\sc ix}: 5.85), 193\,\AA\ (Fe\,{\sc xii}: 6.2; Fe\,{\sc xxiv}: 7.25), 211\,\AA\ (Fe\,{\sc xiv}: 6.3), and 335\,\AA\ (Fe\,{\sc xvi}: 6.45) \citep[see][]{2010A&A...521A..21O,2012SoPh..275...41B}. These AIA data are obtained from the Joint Science Operations Center\footnote{\url{http://jsoc.stanford.edu}}. The AIA data are processed using \texttt{aia\_prep} procedure distributed in the IDL solarsoft library \citep[][]{1998SoPh..182..497F}. They have an image scale of 0.6\arcsec\,pixel$^{-1}$ and cadence of 12\,s\ (during the impulsive phase, the cadence of AIA 131\,\AA\ images varied from 8.22\,s to 15.76\,s, with an average cadence of 12\,s). 

To evaluate thermal characteristics of the flaring region, we complement AIA data with near co-temporal images obtained from the X-Ray Telescope \citep[XRT;][]{2007SoPh..243...63G} onboard Hinode \citep[][]{2007SoPh..243....3K}. In particular, we use XRT image sequences recorded by its thin, medium and thick Be filters between UT\,12:47 to UT\,13:15. These XRT data are retrieved from Hinode Science Data Centre Europe\footnote{\url{http://sdc.uio.no/sdc/}}. The XRT data are processed using \texttt{xrt\_prep} procedure available in IDL/solarsoft. The data have cadence of about 35\,s and an image scale of approximately 1\arcsec\,pixel$^{-1}$. 

The Solar Terrestrial Relation Observatory Ahead spacecraft \citep[STEREO-A;][]{2008SSRv..136....5K} orbiting the Sun was at an Earth Ecliptic longitude of about $-58$\textdegree\ at time of flare. Hence the EUV Imager \citep[EUVI;][]{2008SSRv..136...67H} on STEREO-A had an on-disk view of the flaring region and its surroundings. The STEREO data are retrieved from the Virtual Solar Observatory\footnote{\url{https://sdac.virtualsolar.org/cgi/search}}. In particular, we considered EUVI 171\,\AA\ image sequence. We processed these STEREO data using \texttt{secchi\_prep} procedure available in IDL/solarsoft. The data have an image scale of 1.59\arcsec\,pixel$^{-1}$ and cadence of 75\,s.

Furthermore, the impulsive phase of the flare is also observed by the Fermi Gamma-ray Burst Monitor \citep[GBM;][]{2009ApJ...702..791M}. The NaI detectors on GBM are sensitive to an energy range from a few keV to about 1\,MeV and can detect hard X-ray emission from solar flares. We computed Fermi GBM photon count rates in four energy bands, namely, 10-14\,keV, 14-25\,keV, 25-50\,keV, and 50-100\,keV, by combining signal from the three most sunward detectors minus the three least sunward detectors. We used OSPEX package for Fermi data analysis. We also complemented these Fermi/GBM data with the soft X-ray flux in the 1-8\,\AA\ band recorded by the GOES.

\begin{figure*}
\begin{center}
\includegraphics[width=0.75\textwidth]{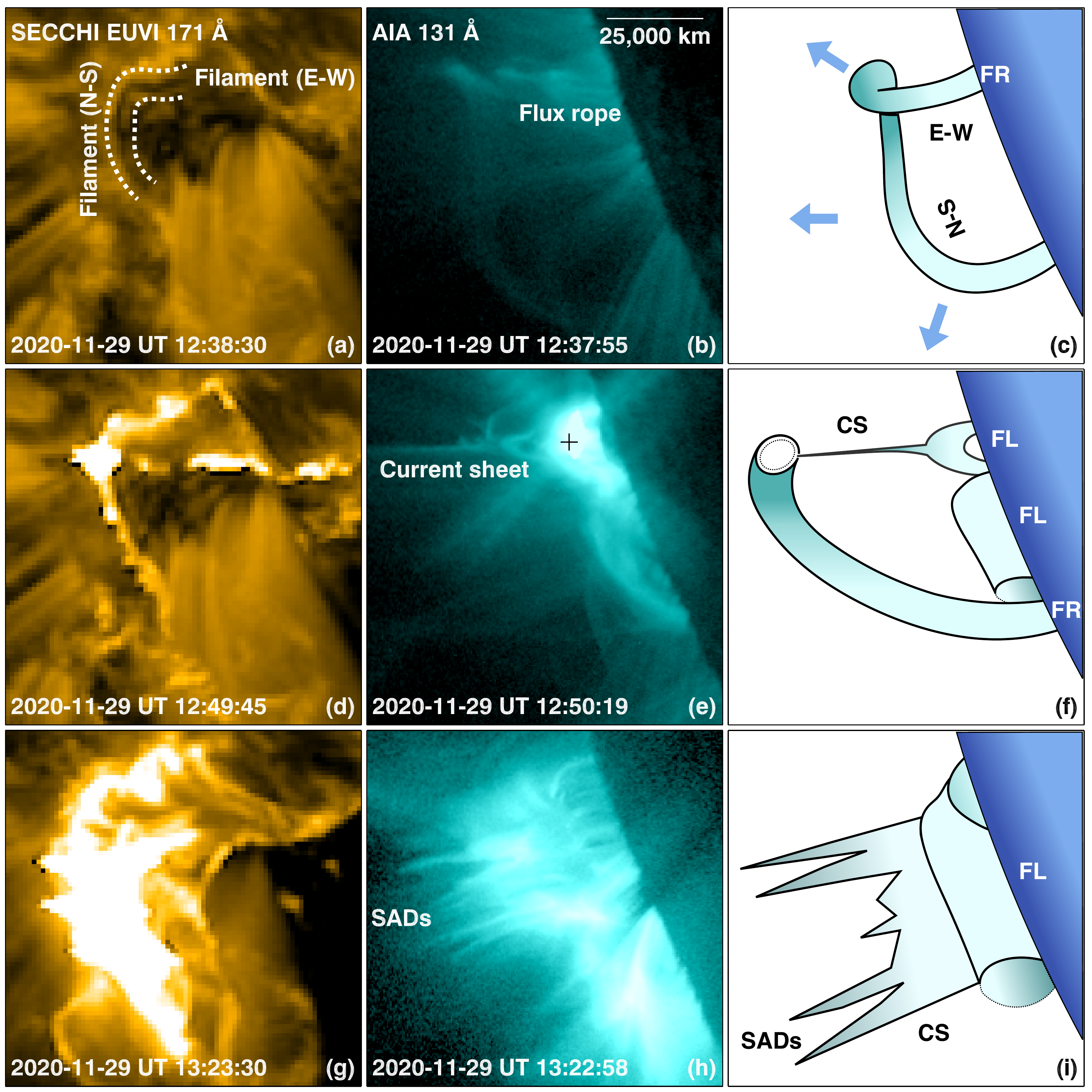}
\caption{Solar flare from different perspectives. The image sequence in the left column displays the flare as seen from above with the EUVI 171\,\AA\ filter on the sun-bound STEREO-A spacecraft. Solar north is up. The three time-steps illustrate the flaring region during the pre-flare phase (top), impulsive rise phase (middle) and main phase (bottom). The SDO limb is about 2\textdegree\ to the west of the western edge of the displayed STEREO field of view. Middle column shows flare evolution from the side as seen with the AIA 131\,\AA\ filter on the earth-bound SDO spacecraft. The hot flux rope, current sheet and supra-arcade downflows (SADs) are marked. The reference plus symbol in the middle panel marks the flaring loop arcade above the solar limb at $(-923.7\arcsec, -362.7\arcsec)$ with respect to the solar disk center as seen from the SDO. The right column illustrates schematically the evolution of the flare and its structure. In the top right panel, the EW and NS sections of the flux rope (FR) are shown. In the middle right panel, flaring loops (FL), initial current sheet (CS) over the E-W segment of the filament and an erupting flux rope (FR) that enters into the plane from behind the current sheet are illustrated. Only the cross section of FR above the current sheet is shown. In the bottom right panel, flaring loops (FL), current sheet in the main phase (CS) that is corrugated by the associated in-plane supra-arcade downflows (SADs) are illustrated. See Sect.\,\ref{sec:obs} for details. Animation of left and middle columns (covering time period UT\,12:30-13:25) is available online.\label{fig:illustration}}
\end{center}
\end{figure*}

\section{2020 November 29 M-class flare\label{sec:event}}

\begin{figure*}
\begin{center}
\includegraphics[width=0.75\textwidth]{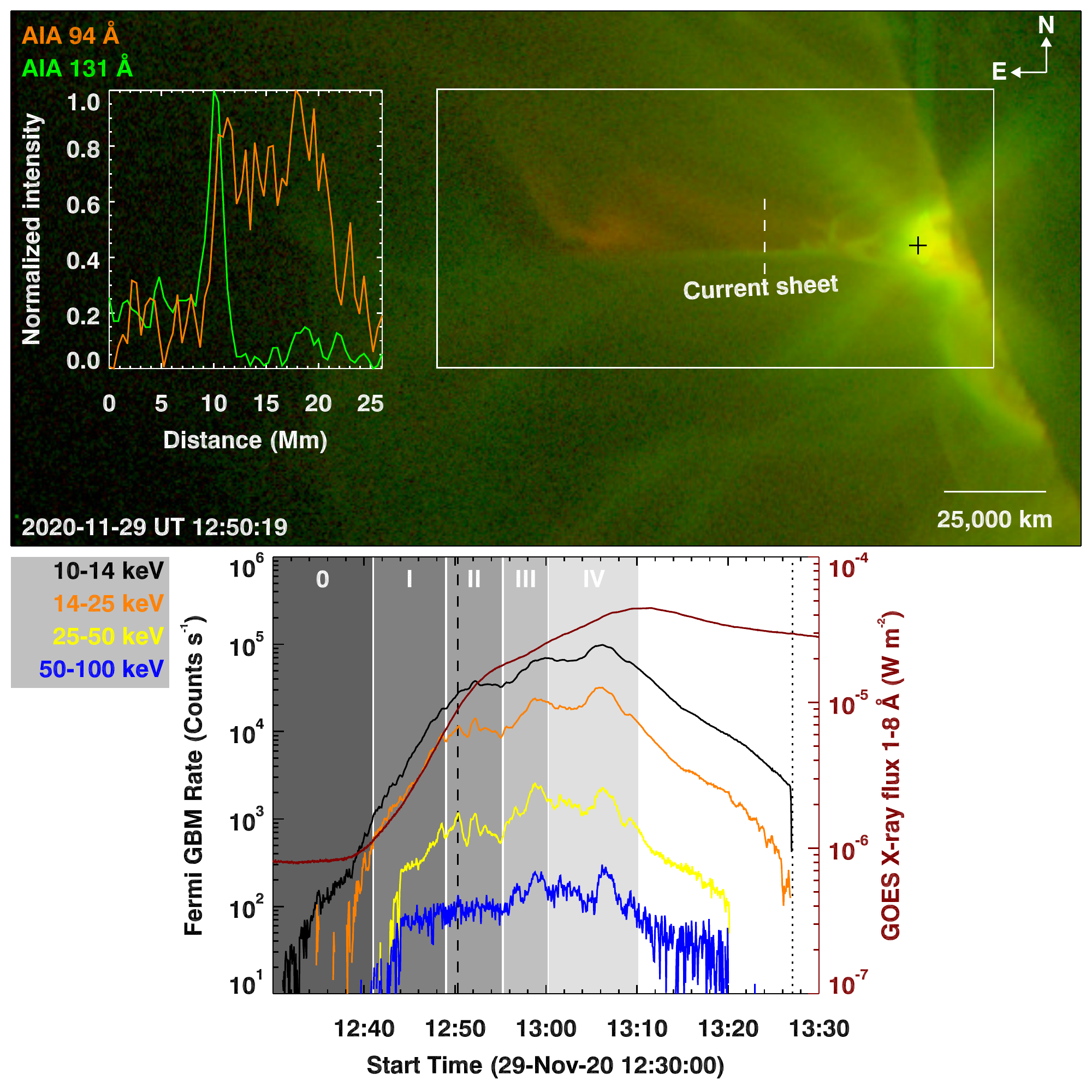}
\caption{The M-class solar flare on 2020 November 29. Top panel displays a composite map of the flaring region at the south-eastern limb of the Sun. The orange and green shaded features identify the EUV emission recorded by the SDO/AIA 94\,\AA\ and 131\,\AA\ filters mainly at temperatures of about 7\,MK and 10\,MK, respectively. The white rectangle encompasses the bright loop system, flux rope and the current sheet during the rise phase of the flare. The reference plus symbol has same coordinates as in Fig.\,\ref{fig:illustration}. In the inset, we plot 131\,\AA\ and 94\,\AA\ intensity profiles across the current sheet and flux rope structures at the location of the vertical dashed line. Bottom panel is a time series of the count rates of hard X-ray photons recorded by the Fermi/GBM (black: 10-14\,keV; orange: 14-25\,keV; yellow: 25-50\,keV; blue: 50-100\,keV). Soft X-ray flux in the 1-8\,\AA\ band recorded by the GOES is plotted in brown. The five grey bands (numbered 0-IV) mark the time periods of different stages of the current sheet evolution that we observed. The dashed vertical line around UT\,12:50 marks the time-stamp of the images in the top panel. The dotted line around UT\,13:27 marks the beginning of Fermi night. See Sect.\,\ref{sec:obs} for details. An annotated animation of this figure is available online. A time series of the top panel (covering time period UT\,12:30-13:25, but without the inset) along with a running time slider is shown in the movie. When the slider enters a particular phase, a close up view of the rectangular region from the top panel appears in the right along with a description of key events during that phase. In particular, when the slider crosses each of phase-I through phase-IV, respective animations of these phases as displayed in Figs.\,\ref{fig:p12} and \ref{fig:p34} also appear on the right side with more description and timing information. The view is played back at five frames per second. \label{fig:overview}}
\end{center}
\end{figure*}

Using the STEREO/EUVI 171\,\AA\ filter, we detected a $\Gamma$-shaped filament at the footpoints of the flaring loop arcade seen with the SDO/AIA. From the vantage point of sun-bound STEREO-A, a part of the filament is oriented roughly in the east-west (E-W) direction, whereas the rest makes a 90\textdegree\ turn along the north-south (N-S) direction, with a NE corner\footnote{Based on the Carrington longitudes, we found that the NE corner of the filament seen in the STEREO images is about 9\textdegree\ behind the solar limb as seen with the SDO.}. We therefore label the east-west part of the flux rope, current sheet and flare arcade as E-W, and the north-south parts as N-S (Fig.\,\ref{fig:illustration}a). Above the E-W filament segment, AIA 131\,\AA\ images revealed an elongated and twisted high-temperature structure \citep[e.g.][]{2011ApJ...732L..25C,2012NatCo...3..747Z,2013ApJ...769L..25C,2013ApJ...764..125P}. Thus, the structure is also recorded by the AIA 94\,\AA\ filter that responds to hot plasma around 7\,MK. Similarly, XRT Be filters also reveal this feature. But it is not seen in other AIA channels that are sensitive to emission from cooler plasma. This is also the feature that expands outwards and erupts. Furthermore, its spatial morphology (elongated and twisted) is different from the flaring loop arcade that forms at latter stages of the flare. All these characteristics suggest that the feature is a hot flux rope at coronal heights (Fig.\,\ref{fig:illustration}b). A larger AIA field of view covering the flaring region is displayed in the top panel of Fig.\,\ref{fig:overview}. The time series of soft X-ray flux and hard X-ray count rates are plotted in the lower panel of Fig.\,\ref{fig:overview}. Based on these light curves, the start of increase in hard X-rays around UT\,12:39 marks the onset of flare, and the rise phase (rise in soft x-rays) lasts until UT\,13:10 and so does the impulsive phase (hard X-rays). After this period, the main phase of the flare starts.

The key events associated with the flare as seen with STEREO are as follows. There is weak or little filament activity until UT\,12:39. First along the E-W filament, bright knots appear around UT\,12:39. Then the bright knots grow to form ribbons starting around UT\,12:46 (bright knots and flare ribbons are shown in Fig.\,\ref{fig:illustration}d). These E-W ribbons separate through UT\,13:10. Along the N-S filament, bright knots appear around UT\,12:46. These knots grow to form ribbons starting around UT\,12:58 and the separation of these flare ribbons starts around UT\,13:01. After around UT\,13:10, flaring activity is mostly along the N-S filament channel (Fig.\,\ref{fig:illustration}g; see online animation associated with Fig.\,\ref{fig:illustration}). 

Similarly, the key events associated with the flare as seen with AIA are as follows. The slow rise of flux rope starts around UT\,12:39. Then the flare activity first develops over the E-W filament. The onset of accelerated expansion of the flux rope is observed around UT\,12:44. Then a thin linear feature appears over the flaring arcade around UT\,12:48 (Fig.\,\ref{fig:illustration}e). Over N-S filament, supra-arcade downflows (SADs) become noticeable around UT\,13:06 and become clearer after that (Fig.\,\ref{fig:illustration}h; see online animation associated with Fig.\,\ref{fig:illustration}).

The initial flux rope is illustrated in Fig.\,\ref{fig:illustration}c. Given that the E-W filament segment, over which the flaring activity is first detected, lies along the line of sight, we initially observed face-on flaring loops. The N-S segment of the filament lies in the plane of the sky when viewed from the SDO vantage point. This resulted in a side-view of the loop arcade when the flaring activity propagated along the N-S filament segment. The expanding flux rope and flaring loop arcades over the E-W and N-S filament segments are illustrated in Figs.\,\ref{fig:illustration}(f) and (i).

Based on STEREO images that displayed ribbons, and AIA observations that revealed overlying flare arcade, it is clear that the event is a classical two-ribbon flare. Solar flare models predict that a current sheet forms at the wake of an erupting flux rope. The observed SAD activity after UT\,13:10 (i.e. during the main phase of the flare) over the N-S filament segment suggests the presence of an apparent face-on current sheet overlying the flaring arcade (as illustrated in Fig.\,\ref{fig:illustration}i). Given that the flare is first detected over the E-W filament segment, a current sheet is expected also over the initial E-W flare arcade, that was predominant during the impulsive rise phase of the flare. To this end, an elongated linear feature is detected directly over the E-W flare arcade at the wake of erupting flux rope around UT\,12:48. 

The elongated structure is clearly visible in the AIA 131\,\AA\ images (Fig.\,\ref{fig:overview}). Sections of this feature are also sporadically seen in the AIA 94\,\AA\ images (Fig.\,\ref{fig:aiamap}). Adjacent to the elongated feature, the leg or flank of the erupting flux rope is detected in the AIA 94\,\AA\ images (Figs.\,\ref{fig:overview}, Appendix\,\ref{sec:app}, and \ref{fig:aiamap}). The elongated feature that is apparently thinner (seen primarily in the 131\,\AA\ images) and the more diffused flux rope (detected mainly in the 94\,\AA\ images) are distinct structures that are spatially offset in the line-of-sight. The intensity profiles across these two features confirm that the width of the elongated feature in 131\,\AA\ images is clearly much narrower compared to the broader and diffused flux rope leg (see inset in the top panel of Fig.\,\ref{fig:overview}). Furthermore, the diffused flux rope leg is northward with respect to the thin elongated feature.

\begin{figure*}
\begin{center}
\includegraphics[width=\textwidth]{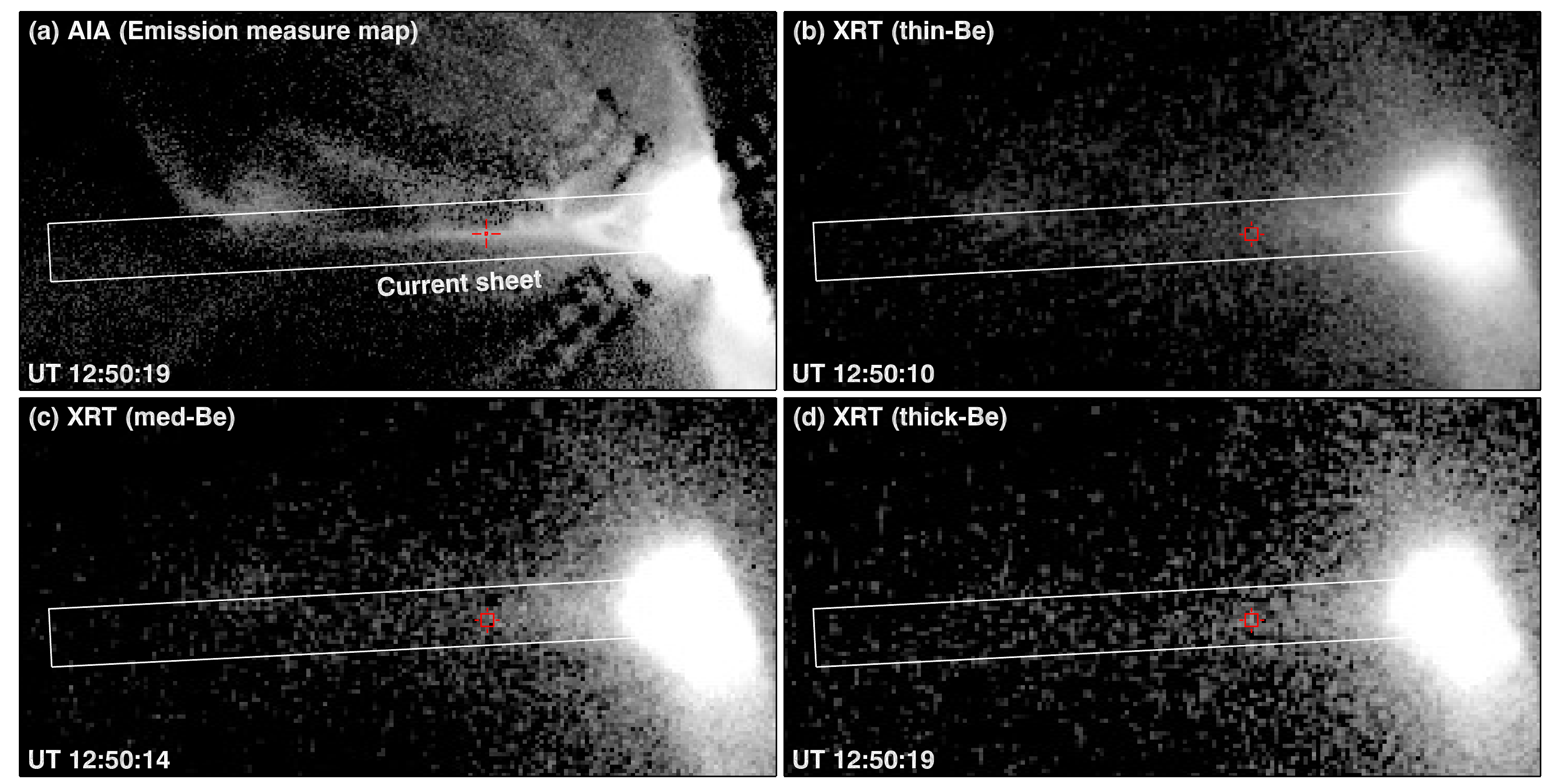}
\includegraphics[width=0.49\textwidth]{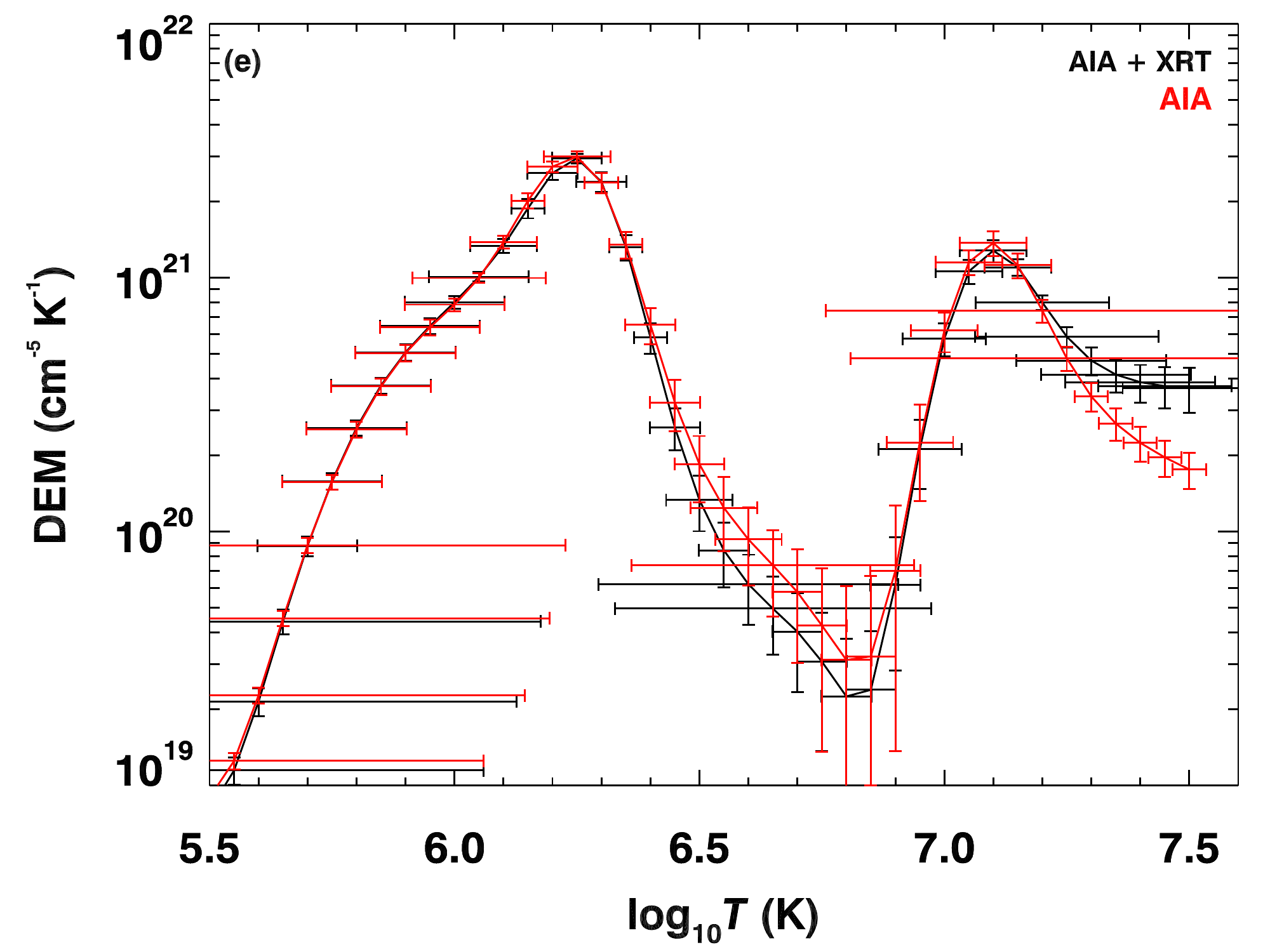}
\caption{Emission characteristics of the current sheet. Panel (a) is the map of emission measure integrated over 9-11\,MK, derived from AIA differential emission measure (DEM) analysis (plotted in log-scale; saturated at $10^{27}-10^{30}$\,cm$^{-5}$). The slanted rectangle highlights the current sheet. Panels (b)-(d) show images obtained with the thin, medium and thick Be filters on Hinode/XRT (plotted in log-scale). In panel (e) we plot DEM as a function of temperature obtained at the location of red symbols in the top panels. Combined AIA (six EUV filters) and XRT (three Be filters; intensity averaged over 3$\times$3 XRT pixels; red symbols in the top panels) DEM curve is shown in black. The red curve is the DEM obtained only with six AIA EUV filters. The horizontal and vertical bars represent the energy resolution of the regularized inversion technique and the 1$\sigma$ errors in DEMs, respectively. See Sect.\,\ref{sec:event} for details. \label{fig:dem}}
\end{center}
\end{figure*}

Here we focus on the thermal characteristics of the thin elongated feature. To this end, we computed the differential emission measure (DEM) with a regularized inversion technique \citep[][]{2012A&A...539A.146H}. Using the images from the six EUV filters\footnote{The AIA images are deconvolved with the filter-dependant instrumental point spread function to remove apparent diffraction patters at that time.}, we computed the DEM over a temperature ($T$) range of log$_{10}T$\,(K) 5.5-7.5. The resulting emission measure map integrated over 9-11\,MK shows the thin elongated feature and the adjacent flux rope (Fig.\,\ref{fig:dem}a). The AIA DEM as a function of temperature from the elongated feature is plotted in Fig.\,\ref{fig:dem}e (red curve). It shows two distinct peaks, one at log$_{10}(T)$\,K$=6.2$, and the second around log$_{10}(T)$\,K$=7$. The first DEM peak corresponds to the cooler coronal structures in the line of sight (detected in other EUV filters of AIA that respond to emission from cooler plasma), whereas the peak at 10\,MK is related to the elongated feature. We further validate this high-temperature nature of the elongated feature using the X-ray images recorded by XRT/Be filters, whose thermal responses peak well over 1\,MK (Figs.\,\ref{fig:dem}b-d). The XRT/thick-Be filter that is sensitive to hot plasma over 10\,MK shows elongated feature that is aligned with the structure seen in AIA 131\,\AA\ images. The other two XRT/Be also show similar elongated features along with the adjacent flux rope. A combined DEM analysis with AIA and XRT data is consistent with the analysis done with AIA EUV filters alone\footnote{The same emission model used to calculate AIA response functions is also used to compute XRT filter responses with CHIANTI atomic database version 9 \citep[][]{1997A&AS..125..149D,2019ApJS..241...22D}, assuming coronal abundances \citep[][]{1992PhyS...46..202F}.}. This combined DEM curve (Fig.\,\ref{fig:dem}e; black) also reveals a hot 10\,MK emission component at the elongated feature.

The elongated feature seen in 131\,\AA\ images has three main characteristics. It is visible only at the wake of an erupting flux rope over the E-W flaring arcade. It is composed of 10\,MK hot plasma  \citep[that is consistent with previous observations, e.g.][]{2018ApJ...854..122W,2018ApJ...866...64C}. It is distinction from the adjacent, diffused leg of the erupting flux rope (better seen in 94\,\AA\ images). Based on these characteristics, the elongated feature we observed is a strong candidate for an edge-on current sheet over the face-on E-W flare arcade (as illustrated in Fig.\,\ref{fig:illustration}f).

When seen from the SDO vantage point, there is a clear separation (spatial offset) between the flaring activity over these E-W and N-S sections of the filament channel. The core of the flux rope and the flaring arcade are first seen over the E-W part of the filament. The thin vertical structure that we suggest to be a current sheet is also seen over this E-W flaring arcade. This enabled us to study a current sheet with both an edge-on and face-on view in the same flare (see online movie associated with Figs.\,\ref{fig:illustration} and \ref{fig:overview}). 

\section{Current sheet evolution\label{sec:cs}}

Here our focus is on the impulsive phase of the flare and the rarely observed dynamics of the associated initial current sheet. Thus, in the rest of the paper, we mainly discuss the evolution of the current sheet with the edge-on view (from Earth) that overlies the E-W segment of the filament.

\subsection{Phase-0 (preflare phase: UT\,12:30-12:41)\label{sec:p0}}
During the pre-flare phase, we observe some activity beneath the E-W portion of the filament. Overlying this region, AIA 131\,\AA\ filter detected a hot flux rope like feature at the south-eastern limb (see top row in Fig.\,\ref{fig:illustration})

Below the flux rope there is no clear indication of a current sheet. The rise of the flux rope (both its E-W and N-S parts) could perhaps be initiated by nonequilibrium of the flux-rope or weak reconnection at lower altitudes below the filament (filament seen in STEREO-A; images after the flare show that the filament did not completely erupt). As the flux rope slowly rises, it drags along with it and stretches out the low-lying loops. At this stage, the soft X-ray flux detected by GOES is at a background level with a slow increase towards the end of the phase; the background is set by an earlier on-disk C-class flare in AR 12786 (whose peak is around 11:44 UT). The count rates in the Fermi lower energy band covering 10-14\,keV count rate shows a steady increase, indicating increasing levels of flux rope activity/heating.

\subsection{Phase-I (formation of current sheet: UT\,12:41-12:49)\label{sec:p1}}

\begin{figure*}
\begin{center}
\includegraphics[width=0.49\textwidth]{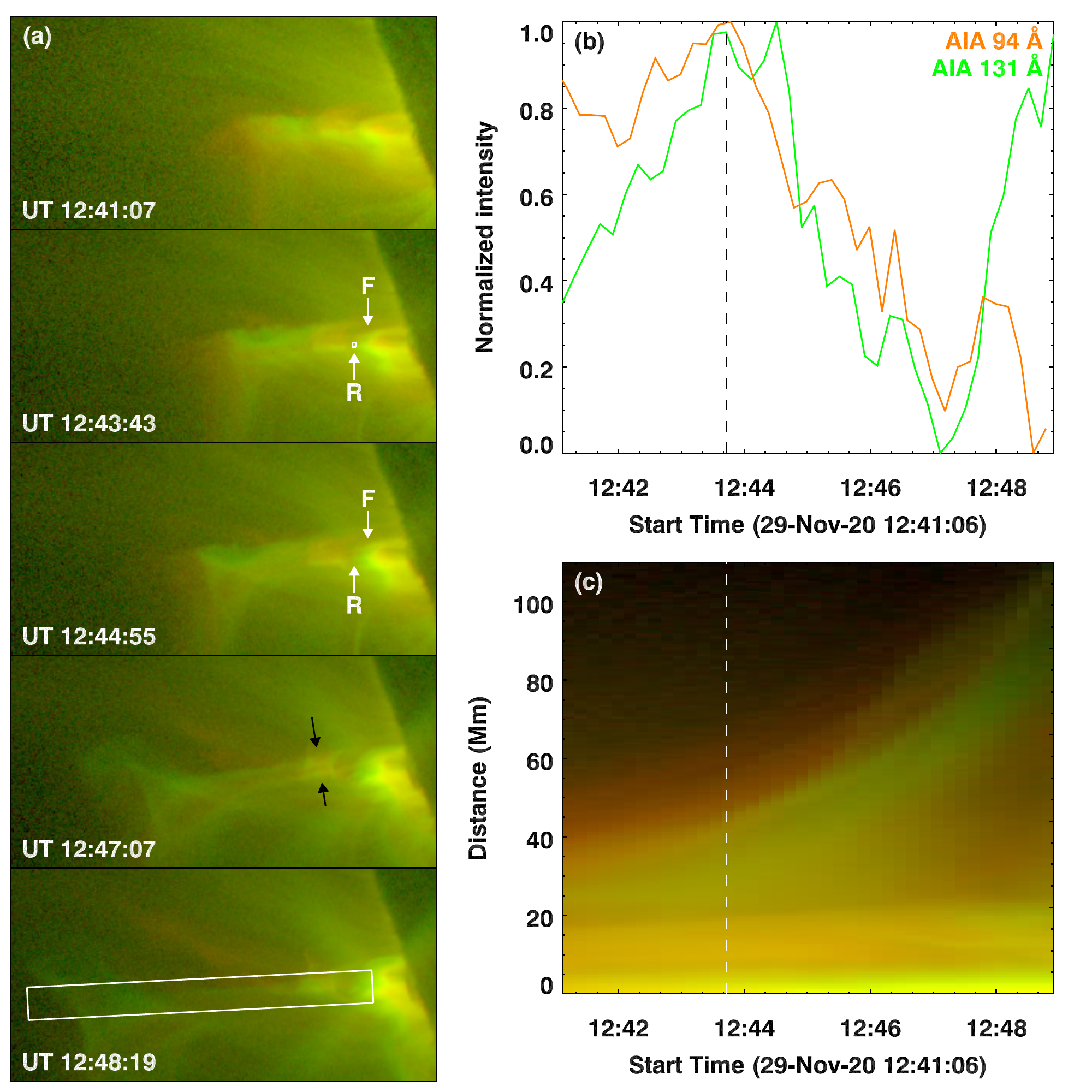}
\includegraphics[width=0.49\textwidth]{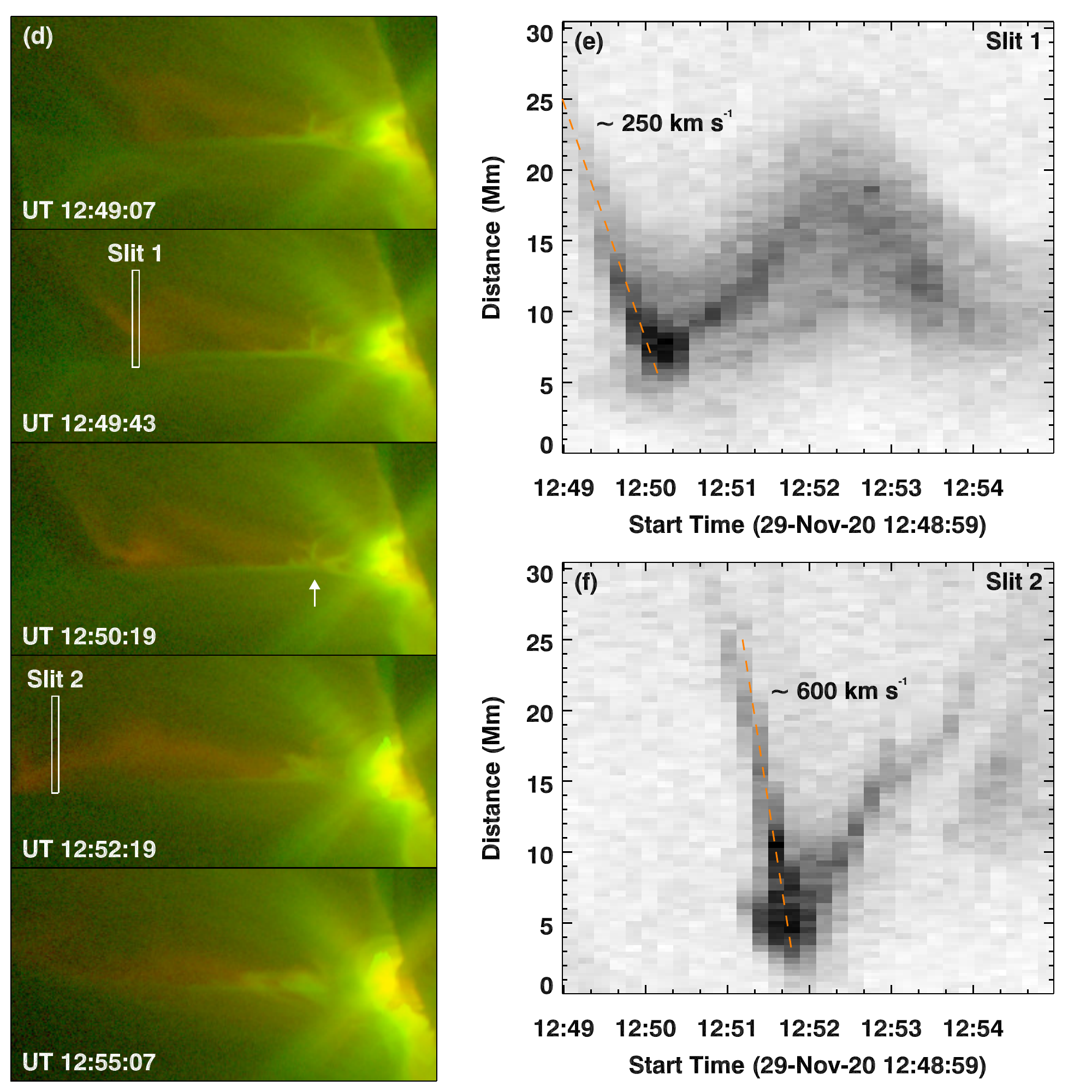}
\caption{Expansion of the flux rope and formation of the current sheet (phases I and II). Panel (a) shows a closer view of the region outlined with the rectangle in Fig\,\ref{fig:overview}. The green and orange shades have the same meaning as in Fig.\,\ref{fig:overview}. Time runs from top to bottom. The white arrow points to the location R (the white square) where reconnection is likely to be initiated. It is located just above the first flare loop (F). The black arrows indicate the location where the legs of the expanding flux rope interact. The slanted rectangular slit in the lower panel marks the elongated feature that developed in the wake of the erupting flux rope, which we suggest is a current sheet. The EUV emission light curves from the location R of reconnection are plotted in panel (b). The integrated emission across the slanted slit is plotted along its length as a function of time in panel (c). In panels (b) and (c), the dashed vertical lines mark the time-stamp of the start of reconnection as seen in the AIA images. Panel (d) is same as panel (a), but shows the development of the elongated current sheet after the initiation of reconnection. The two slits (1 and 2) identify locations where the flank of the flux rope interacts with the current sheet. The white arrow points to the western, sunward tip of the current sheet. In panels (e)-(f), time-distance maps derived using AIA 94\,\AA\ emission from slits 1 and 2 are plotted (in an inverted color-scheme). Approximate slope of the orange dashed lines in both cases is quoted in units of speed (km\,s$^{-1}$). See Sects.\,\ref{sec:p1} and \ref{sec:p2} for details. \label{fig:p12}}
\end{center}
\end{figure*}

The onset of phase-I, the start of the flare, is marked by several simultaneous features. First, a small dark void appears in the EUV images (marked by white arrow, letter R and box in Fig.\,\ref{fig:p12}(a) at UT\,12:43:43). secondly, the AIA light curves exhibit a rapid drop in intensity at that void location caused by the separation of the overlying flux rope and underlying flare arcade. Thirdly, the outward motion of the flux rope accelerates as the void rapidly grows longer (Figs.\,\ref{fig:p12}(b)-(c)). Based on these simultaneous features, we suggest that void is the location where reconnection is initiated. It could possibly be a neutral point, separator or quasi-separator, which are all prone to reconnection \citep[][]{2000priest}. 

The accelerating and erupting flux rope is associated with elongated features in its wake, but we still do not have a clear indication of a current sheet. Around UT\,12:47:07, these elongated features begin to merge (black arrows in Fig.\,\ref{fig:p12}). These elongated features and their merging is consistent with the scenario of magnetic reconnection at the wake of erupting flux rope. The first indication of current sheet becomes apparent around UT\,12:48:19. Thus in this case the current sheet becomes clear nearly 5\,minutes after the first indications of reconnection beneath the flux rope. Once formed, the linear current sheet is the dominant emission feature between the erupting flux rope and the flare arcade with little line of sight confusion. With the appearance of the elongated current-sheet like structure, phase-I ends. Around the same time, STEREO-A images show bright flare knots, which grow into weak flare ribbons developing on the disk, mainly along the E-W part of the underlying filament (see movie and middle row of Fig.\,\ref{fig:illustration}).

Both the preflare conditions and the expansion of flux ropes have been well documented in the literature \citep[see more recent investigations on the expansion of flux ropes by][]{2018ApJ...853L..18Y,2019SciA....5.7004G,2020ApJ...894...85C}. Nevertheless, pinning down the location of reconnection beneath the expanding flux rope and the aforementioned several simultaneous features including the onset of current sheet formation have not been observed before with this level of clarity.

\subsection{Phase-II (quasi-stable evolution of current sheet: UT\,12:49-12:55)\label{sec:p2}}

The elongated structure now is clearly seen and it is further stretched as the flux rope continues to accelerate away from the Sun. The STEREO images show that the E-W flare ribbons are fully formed and that they brighten and begin to separate. At the sunward, western tip of the current sheet there are cusp-shaped loops (white arrow in Fig.\,\ref{fig:p12}d). At the eastern tip of the current sheet, one flank of the erupting flux rope exhibits a whiplash motion, clearly seen in the AIA 94\,\AA\ image sequence (orange shaded features in Fig.\,\ref{fig:p12}d). A portion of this flux rope flank suddenly sways toward the now clear hot current sheet (see also Fig.\,\ref{fig:dem}) and interacts with it, first at slit-1 around UT\,12:49:07. The time distance map of this swaying motion seen in the 94\,\AA\ filter image sequence at slit-1 is displayed in Fig.\,\ref{fig:p12}e. The material approaches the current sheet at speeds of about 250\,km\,s$^{-1}$. This interaction as seen in AIA 94\,\AA\ images is further discussed in the Appendix\,\ref{sec:app} and illustrated in Fig.\,\ref{fig:p2_94}. The flank appears to sway away from the current sheet soon after this interaction. As it sways away, that section of the flux rope develops a kink feature after its interaction with the current sheet (see Figs.\,\ref{fig:kink_94}(a)-(b)) in Appendix\,\ref{sec:app}. After 90\,s or so, the flux rope flank interacts with the current sheet again at slit-2 (UT\,12:51:55). The material approaches the current sheet faster than during the first interaction, at speeds of 600\,km\,s$^{-1}$ and retracts away from it soon after the interaction. Similar to the events after the first interaction at slit-1, that section of the flux rope interacting with the current sheet at slit-2 also develops a kink feature (see Figs.\,\ref{fig:kink_94}(c)-(d) in Appendix\,\ref{sec:app}). Although the section of the flux rope bounded by slits-1 and -2 also approaches the current sheet, it does not develop similar kink features after swaying away. Thus the nature of interaction of the flux rope with the current sheet at slits-1 and -2 is likely episodic. These swaying motions of the flux rope occur simultaneously with the oscillations of adjacent coronal loops to its north, seen in the AIA 211\,\AA\ images (see discussion in Appendix\,\ref{sec:app}). 

Afterward, the east-most tip of the flux rope exhibits a sliding motion as it interacts with the current sheet and gets deflected. At the sunward western tip the cusp shaped loops turn more chaotic during this period (see online animation associated with Fig.\,\ref{fig:overview}). Throughout this phase, even though current sheet is perturbed from both ends, its structure is preserved and exhibits a quasi-stable evolution.

\begin{figure*}
\begin{center}
\includegraphics[width=0.49\textwidth]{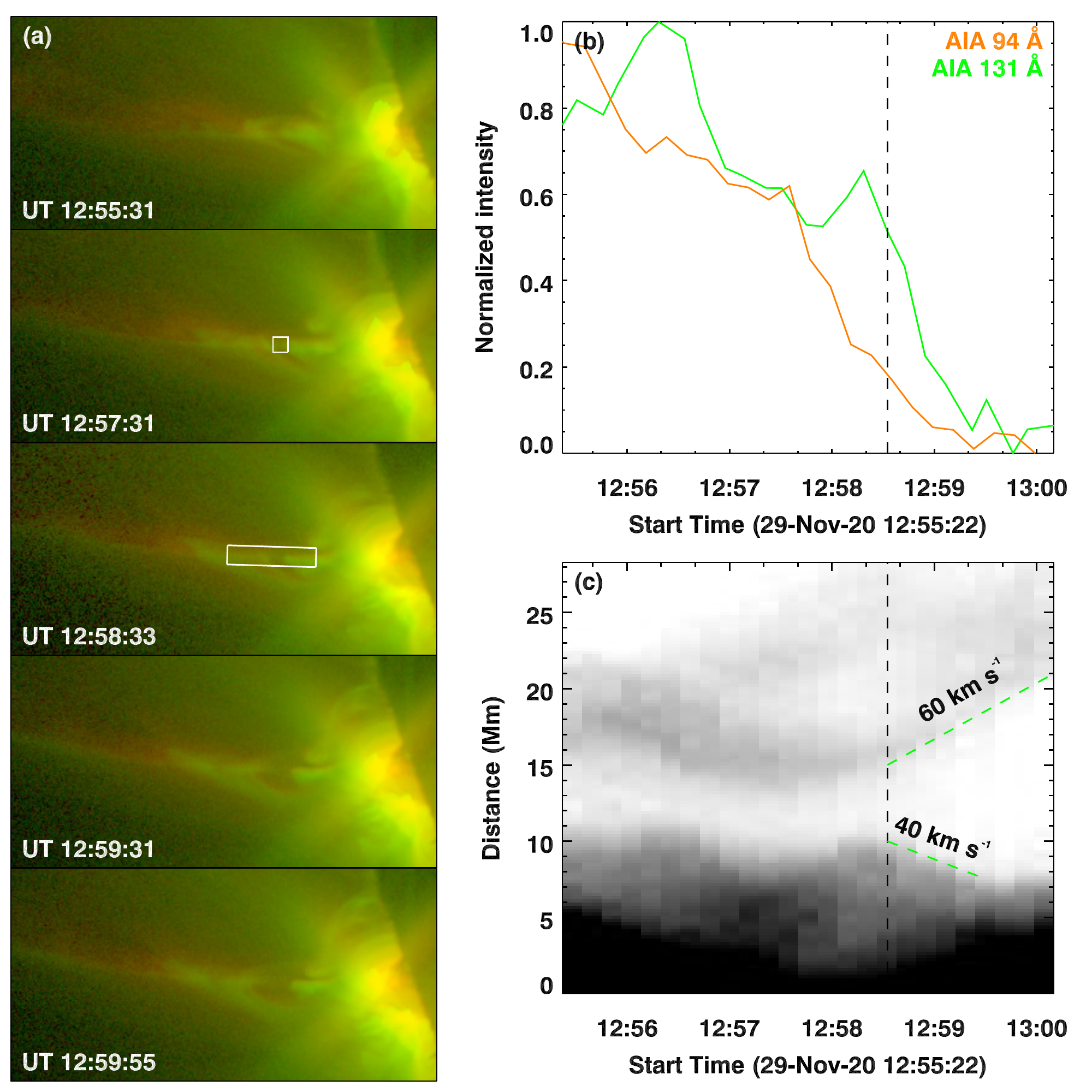}
\includegraphics[width=0.49\textwidth]{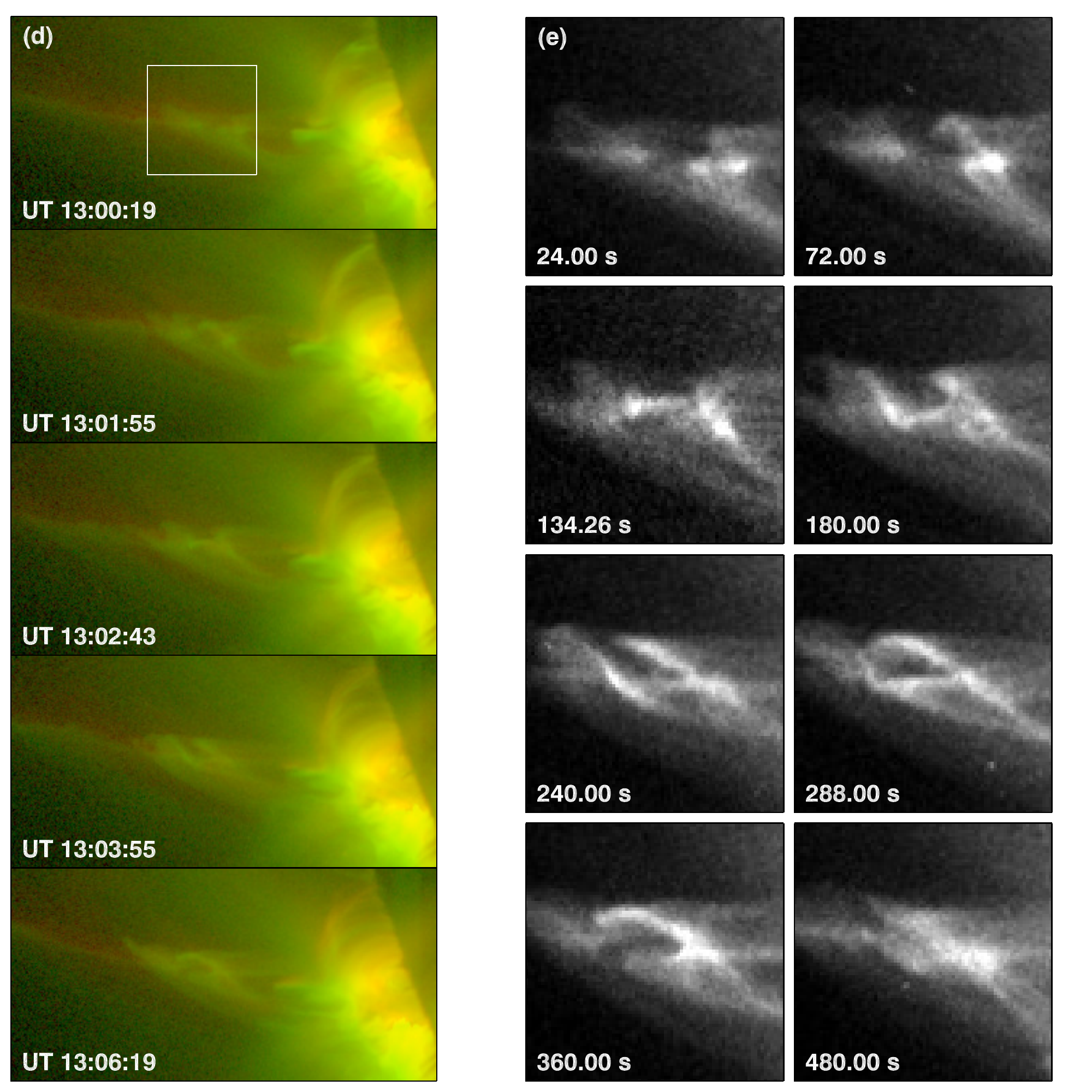}
\caption{Disruption of the current sheet (phases III and IV). The format of panels (a)-(c) is same as in Figs.\,\ref{fig:p12}(a)-(c). Panel (a) shows a time sequence of disturbances along the current sheet that developed and grew from its sunward western tip (white arrow in Fig.\,\ref{fig:p12}). The light curves in panel (b) and computed from the white square in panel (a). The AIA 131\,\AA\ time-distance map (panel c; in an inverted color-scheme) is derived from the slanted rectangle in panel (a). Here 0\,Mm marks the western side of the rectangle. Slopes of the green dashed lines are quoted in units of speed (km\,s$^{-1}$). Panel (d) shows a time sequence of the disruption of the current sheet. The white squared region marks the field of view shown in panel (e). A base-difference image sequence of AIA 131\,\AA\ emission from the white square in panel (d) is shown in panel (e). Here the base image at UT\,13:00:19 is subtracted from the subsequent time series. The time elapsed (in seconds) after the base image is quoted. See Sects.\,\ref{sec:p3} and \ref{sec:p4} for details.\label{fig:p34}}
\end{center}
\end{figure*}

\subsection{Phase-III (onset of current sheet disruption: UT\,12:55-13:00)\label{sec:p3}}

The previously cusp-shaped loops at the sunward tip of the current sheet suddenly grow in size and further stretch the current sheet. The STEREO images show a continuing brightening and separation of the E-W flare ribbons and formation of the N-S ribbons. In a similar way to phase-I, a void-like feature, that is discernible in both AIA 94\,\AA\ and 131\,\AA\ filter images, appears (Fig.\,\ref{fig:p34}a). This indicates another phase of reconnection of flux rope legs that are still connected to the Sun. Just as the flux rope and loops separate suddenly around UT\,12:58:33, the AIA light curve intensity at this location rapidly decreases (Fig.\,\ref{fig:p34}b). The separation of flux rope and loop system is seen in the time-distance map derived from this region (green dashed lines in Fig.\,\ref{fig:p34}c). This indicates that reconnection is perhaps weakening as the flux rope is now completely detached at that end. This also marks the end of phase-III.

\subsection{Phase-IV (disruption of initial current sheet UT\,13:00-13:10)\label{sec:p4}}

Perturbations at the current-sheet and flare-arcade interface were initiated in phase-II and grew through phase-III. These growing perturbation features could possibly be a manifestation of the superposition of several expanding loops that are associated with the motion of the current sheet to higher altitudes. Nevertheless, there is an apparent interaction between the two systems. These growing perturbation features become more regular and orderly starting around UT\,13:00 (Fig.\,\ref{fig:p34}d). The resulting interaction exhibits growing swirl-like eddies (Fig.\,\ref{fig:p34}e), that are regular and persistent for nearly 10\,minutes until UT\,13:10. The organized swirls mostly weaken or stop around UT\,13:10:33. During this phase, the STEREO images show that the N-S ribbons are fully formed and separate. Also, there is an extra brightening of the E-W ribbons and of one particular flare loop (see the frame at UT\,13:06 in the online movie associated with Fig.\,\ref{fig:illustration}), which suggests a burst of enhanced reconnection that may be associated with a disruption of the current sheet.

During and after this phase, apparent activity of initial current sheet that first appeared at the end of phase-I diminishes and it continually fades away from view in the AIA 131\,\AA\ images, indicating its likely disruption. At the same time, as shown in Fig.\,\ref{fig:illustration}, the flare activity propagates to the N-S segment of the filament. Now an apparently new current sheet appears face-on over the flare arcade above the N-S filament. We observed a continual growth of flare arcade and the appearance of SADs over the N-S filament around the same time that the initial current sheet over the E-W filament fades away. Therefore, a current sheet is always present during this flare (as soon as the ribbons form). Thanks to this spatial offset in the plane-of-sky, the thin linear structure over the E-W flaring arcade, which formed during the impulsive rise phase of the flare (around UT\,12:48) remains clearly distinguishable from the SAD activity over the N-S filament channel (starting around UT\,13:06), which increased during the main phase of the flare after about UT\,13:10 or so. All these phases are summarized also in an annotated animation accompanying Fig.\,\ref{fig:overview} that is available online.

\section{Hard X-ray emission \label{sec:hxr}}
During the rise phase of the flare, Fermi/GBM detected a gradual rise of count rates in the 10-14\,keV and 14-25\,keV energy bands, suggesting plasma heating. The count rates of the higher energy bands (i.e. 25-50\,keV and 50-100\,keV) are lower, but their light curves are dominated by fluctuations on timescales of a few minutes (Fig.\,\ref{fig:overview}). Generally, such high energy hard X-ray emission is thought to be produced by the bremsstrahlung of nonthermal electrons. However, it is know that that Fermi data are affected by pileup issue \citep{2009ApJ...702..791M}. In that case, increases in X-ray intensity can be alternatively explained as a dominant thermal plasma source and some pileup to the higher energies. To this end, based on spectral analysis, we confirm that the our observations are not affected by pileup issue. Thus, the high energy hard X-ray emission in this case is mostly produced by nonthermal electrons (detailed spectral analysis will be presented in a future publication).

Here we briefly discuss the timing of fluctuations in the hard X-ray emission during different phases of current sheet evolution presented in Sect.\,\ref{sec:cs}. During the accelerated outward motion of the flux rope in Phase-I (i.e. around UT\,12:43:43), Fermi 25-50\,keV count rate begins to increase. Furthermore, as the current sheet becomes apparent around UT\,12:48:19, a short, but clear peak in the FERMI 25-50\,keV light curve is detected just close to the end of phase-I (see Sect.\,\ref{sec:p1} and Fig.\,\ref{fig:overview}). During the two episodes of flux rope interaction with the current sheet in Phase-II around UT\,12:50:19 and UT\,12:51:55, there are two hard X-ray bursts in the Fermi 25-50\,keV time series. On both occasions, similar peaks are also seen in the 14-25\,keV band, and to a lesser extent also in 50-100\,keV light curves (see Sect.\,\ref{sec:p2} and Fig.\,\ref{fig:overview}). As the void continues to evolve in Phase-III, the count rate in the 50-100\,keV band grows. Furthermore, at the time of sudden separation of flux rope and loops around UT\,12:58:33, a drop in Fermi hard X-ray flux is detected (see Sect.\,\ref{sec:p3} and Fig.\,\ref{fig:overview}). Concurrent with swirl-like motions presented in Phase-IV, a new hard X-ray peak appears. Moreover, with the weakening of these swirls around UT\,13:10:33 and the disruption of the initial current sheet, the hard X-ray emission stops growing further and the main phase of the flare that is dominated by soft X-ray emission begins (see Sect.\,\ref{sec:p4} and Fig.\,\ref{fig:overview}).

However, we do not have spatial information on the source regions of hard X-ray emission. Therefore, any direct role of the initial current sheet dynamics in the observed modulation of the high energy hard X-ray emission cannot be established. 

\section{Discussion and Conclusion \label{sec:disc}}
Hard X-ray emission from accelerated nonthermal electrons during the impulsive phase of the flare can originate from both footpoint and coronal sources \citep[e.g.][]{1994Natur.371..495M,2003ApJ...596L.251S,2008A&ARv..16..155K,2013ApJ...767..168L,2015Sci...350.1238C}. In our case, footpoints of the flare are occulted by the solar disk and so the main source region for the hard X-ray emission is likely the corona. We observed multiple nonthermal hard X-ray impulses during the rise phase of the flare that are clearest in the Fermi/GBM 25-50\,keV time series. Similar impulses are also seen in the evolution of both the lower and higher energy bands. For instance, there are three distinct broad bumps in the Fermi 10-14\,keV light curve, seen between UT\,12:49 and UT\,13:10 (see Fig.\,\ref{fig:overview}). Three similar bumps are also evident in the 14-25\,keV light curve. Simultaneous to these broad bumps below 25\,keV, there are much sharper peaks in the 25-50\,keV time series, along with detectable peaks also in the 50-100\,keV time series. This close correspondence in the hard X-ray time series across various energy bands suggests that each of these episodes has a common origin related to the magnetic reconnection that is powering the whole flare. Earlier studies have invoked plasmoid-like loop ejections from flaring regions \citep[][]{2010ApJ...711.1062N}, reconnection outflows \citep[][]{2013ApJ...767..168L} or termination shocks above the flaring loop arcade \citep[][]{2015Sci...350.1238C} as likely drivers to accelerate particles and produce coronal hard X-ray emission. Our observations reveal a dynamic evolution of the current sheet that could possibly regulate the reconnection itself. Numerical models are required to verify if and how evolving current sheets can modulate particle acceleration in the corona.   

Furthermore, thin and long current sheets are prone to tearing mode or secondary tearing mode instabilities that trigger reconnection by ejecting magnetic islands or plasmoids along the current sheet \citep[][]{2001EP&S...53..473S,2007PhPl...14j0703L,2009PhPl...16k2102B}. In this regard, bright blob like emission features that appear along the current sheet during the flare are often interpreted as signatures of such plasmoids \cite[][]{2015SSRv..194..237L,2019SciA....5.7004G}. However, in our case, though the current sheet has evolved through multiple phases, we have not observed any clear signatures of these bright blobs or plasmoids. Even during phase-II where the current sheet is externally perturbed by its interaction with the fast-moving flank of the erupting flux rope (Fig.\,\ref{fig:p12}), the structure evolves quasi-stably with no indication of plasmoid formation or blob-like ejections. The main disturbances to the current sheet that we noted arise from its interface with the underlying flaring arcade. At the end of phase-II, we noticed some disturbances at this interface which grew through phase-III. They then turn more regular and orderly in phase-IV, after which the current sheet fades away. 

These swirl-like eddies are clearly not plasmoids as they move in only one direction while reconnection models predict bidirectional ejection of plasmoids along the current sheet. Two possible explanations occur to us. Firstly, these swirls could indicate shear-induced instabilities, with the shear developing within the sheet or when the expanding plasma slides along a rather static current sheet and gives rise to a velocity gradient at the interface which could then generate Kelvin–Helmholtz instability. Another explanation is that the initial edge-on current sheet is contoured by turbulent SADs \citep[][]{2013ApJ...766...39M}. However, this latter explanation is unlikely because the SADs are sunward propagating while the eddies we observed exhibit an apparent anti-sunward motion.  Nevertheless, whether the swirl-like motions we observed are due to instabilities or SADs, both possibilities offer new ways to extend models of reconnection in dynamic current sheets and their relevance to particle acceleration in flares. 

Given that the flare happened behind the limb, we do not have a complete understanding of the propagation of the flare itself. The formation of ribbons, starting with initial brightening in knots has been modeled by \citet{2017SoPh..292...25P} due to zipper reconnection, which also describes how a flare propagates from one end of a filament channel to the other: more details on this process await MHD modeling. However, current sheet evolution during the rise phase of the flare is rarely observed and we propose that our observations here shed light on how it might develop. Our observations revealed that a dynamic current sheet during the impulsive phase of a flare evolves through multiple phases from its formation to disruption. A slowly rising flux rope stretches the underlying field and forms a long current sheet, whose reconnection accelerates the flux rope more quickly and initiates the impulsive phase with its hard X-ray emission. The long current sheet then evolves quasi-stably but modulated aperiodically by the perturbations nearby, which do influence the reconnection process within the current sheet. Since the dynamics of the current sheet are likely to be regulated by the eruption itself, so are the onset and progression of reconnection at the current sheet. External perturbations could lead to variations in reconnection rates. Indeed, the macroscopic processes that we observed in the evolution of the current sheet could play a role in the conversion of magnetic energy in solar flares.

\acknowledgments
The authors thank the three anonymous referees for constructive comments that helped us improve the presentation of the manuscript. L.P.C. is grateful to Anne K. Tolbert and Brian R. Dennis (both at NASA/GSFC) for their help with Fermi data analysis. X.C. is funded by NSFC grants 11722325, 11733003, 11790303, 11790300, and the Alexander von Humboldt foundation. SDO data are courtesy of NASA/SDO and the AIA, EVE, and HMI science teams. Hinode is a Japanese mission developed and launched by ISAS/JAXA, collaborating with NAOJ as a domestic partner, NASA and STFC (UK) as international partners. Scientific operation of the Hinode mission is conducted by the Hinode science team organized at ISAS/JAXA. This team mainly consists of scientists from institutes in the partner countries. Support for the post-launch operation is provided by JAXA and NAOJ(Japan), STFC (U.K.), NASA, ESA, and NSC (Norway). We acknowledge the use of the Fermi Solar Flare Observations facility funded by the Fermi GI program \url{http://hesperia.gsfc.nasa.gov/fermi_solar/}. We are grateful to STEREO and GOES teams for making the data publicly available. CHIANTI is a collaborative project involving George Mason University, the University of Michigan (USA), University of Cambridge (UK) and NASA Goddard Space Flight Center (USA). This research has made use of NASA’s Astrophysics Data System.
 
\facilities{Hinode, FERMI(GBM), SDO, STEREO}

\appendix

\section{Current sheet and flux rope structures\label{sec:app}}

\begin{figure*}
\begin{center}
\includegraphics[width=0.8\textwidth]{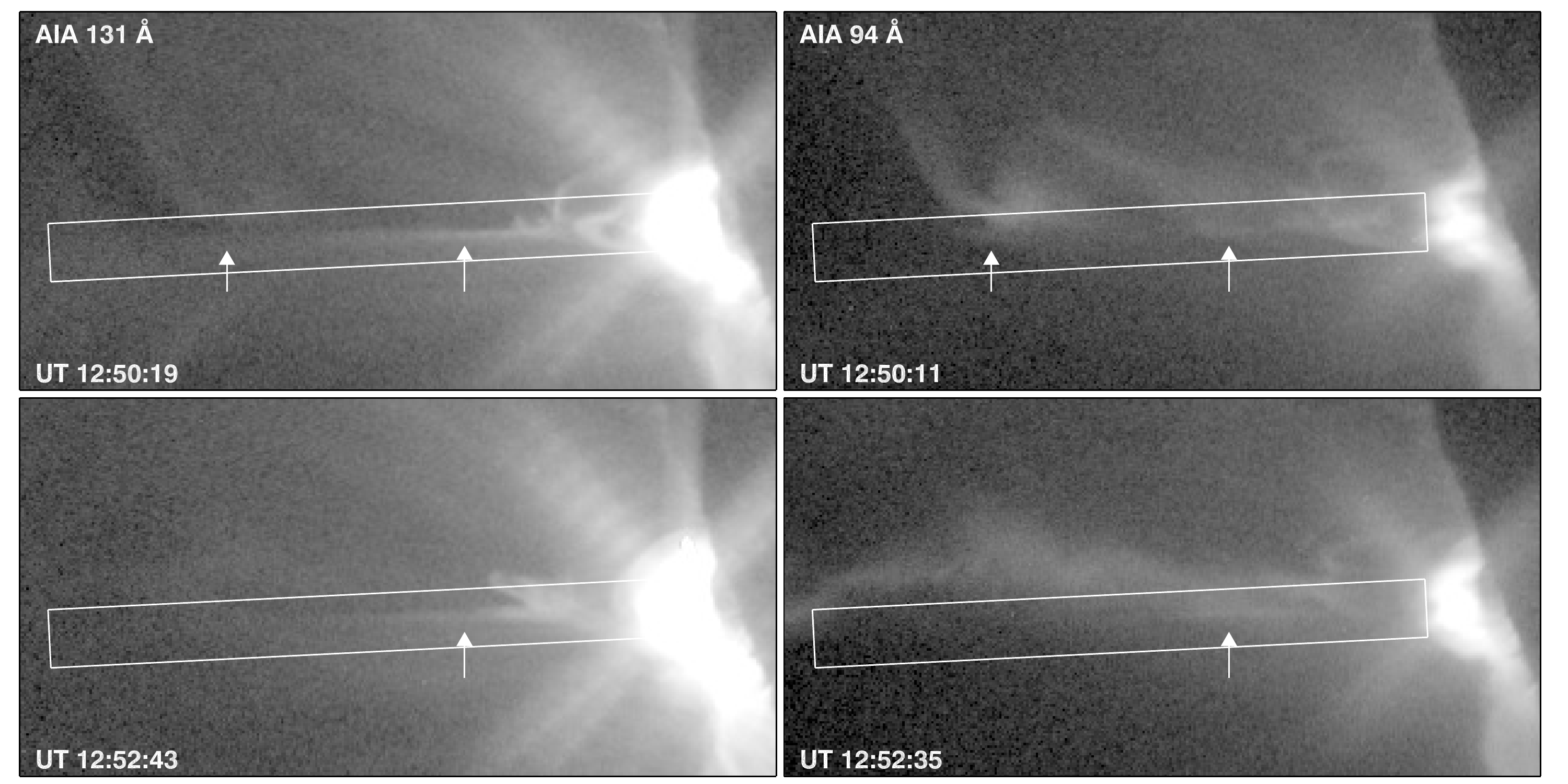}
\caption{Structure of current sheet and adjacent flux rope. The left and right panels show AIA 131\,\AA\ and 94\,\AA\ snapshots at two instances. The slanted rectangle is the same as in Fig.\,\ref{fig:dem}a. In the right panels, the arrows point to sections of thin linear features aligned with the elongated current sheet structure detected in the AIA 131\,\AA\ images. A leg of the erupting flux rope is seen as a diffused structure, on the north side of the slanted box in the right panels. See Sect.\,\ref{sec:event} and Appendix\,\ref{sec:app} for details. \label{fig:aiamap}}
\end{center}
\end{figure*}

\begin{figure*}
\begin{center}
\includegraphics[width=0.49\textwidth]{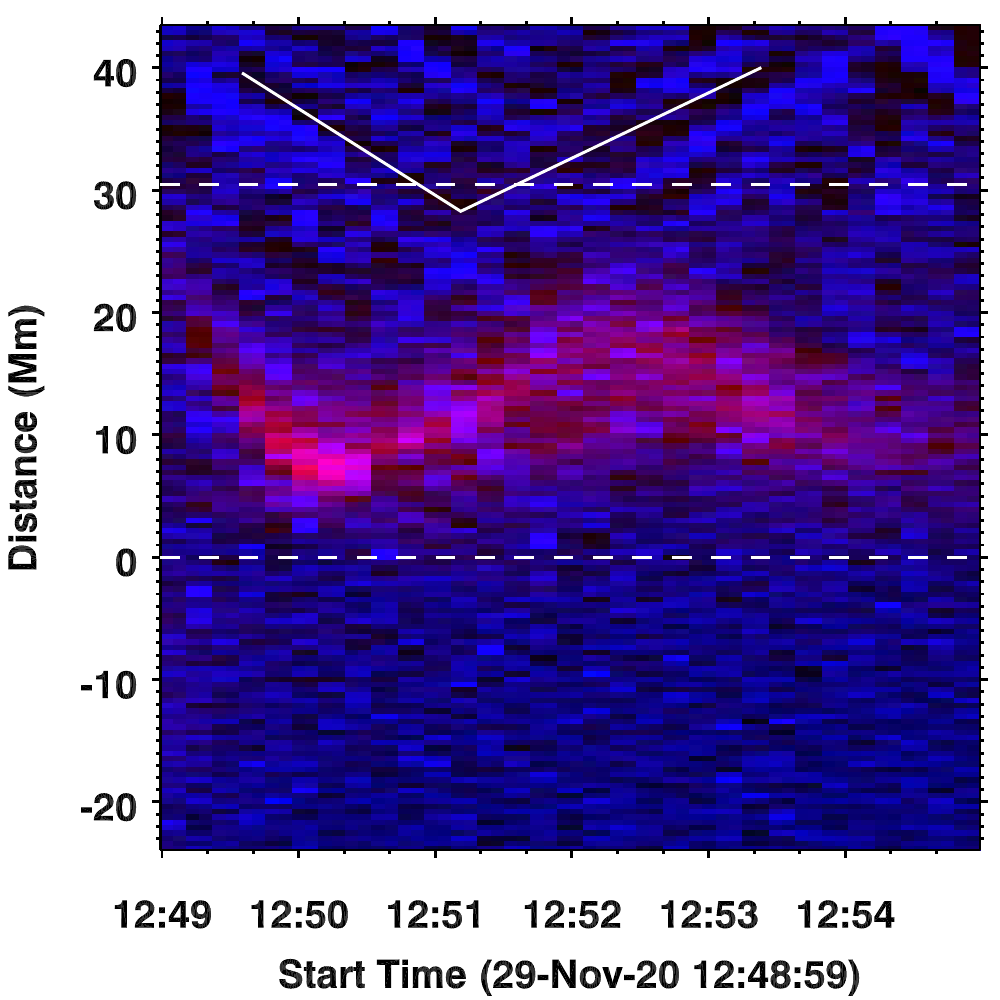}
\caption{Dynamics of the flux rope flank and adjacent loops. Time distance map of the AIA 94\,\AA\ emission (red) and AIA 211\,\AA\ emission (blue) along the entire height of Fig.\,\ref{fig:p12}(d) at slit-1 is shown. The extent of slit-1 in Fig.\,\ref{fig:p12}(d) itself is marked by a pair of horizontal dashed lines. Here 0\,Mm along the vertical axis coincides with the southern tip of slit-1. The red shaded feature is the swaying motion of the flux rope flank seen in the AIA 94\,\AA\ images along slit-1, during the quasi-stable evolution of the current sheet (phase-II; Sect.\,\ref{sec:p2}; Fig.\,\ref{fig:p12}e). The blue V-shaped features on the northern side of the flux rope highlight the swaying motion of adjacent loops seen in the AIA 211\,\AA\ images. See Sect.\,\ref{sec:p2} and Appendix\,\ref{sec:app} for details.\label{fig:p2_94_211}}
\end{center}
\end{figure*}

In the main text, we described the evolution of thin current sheet as seen in the AIA 131\,\AA\ images (see Fig.\,\ref{fig:illustration}). However, sections of the current sheet are sporadically seen also in the AIA 94\,\AA\ images (Fig.\,\ref{fig:aiamap}). This is particularly apparent when the leg of erupting flux rope interacts with the current sheet. The interaction itself is caused by swaying motion of the flux rope. By examining images from other channels of AIA, we found that the flux rope swaying seen in the AIA 94\,\AA\ images is simultaneous with the oscillation of northward coronal loops, clearly seen in the AIA 211\,\AA\ images. As the flux rope expands outward, the adjacent coronal loops are swept into the region previously occupied by the flux rope itself. The back and forth motion of the loops is highlighted in Fig.\,\ref{fig:p2_94_211} (V-shaped solid lines). Such loop motions could be responsible for the swaying motion of the flux rope.

\begin{figure*}
\begin{center}
\includegraphics[width=0.49\textwidth]{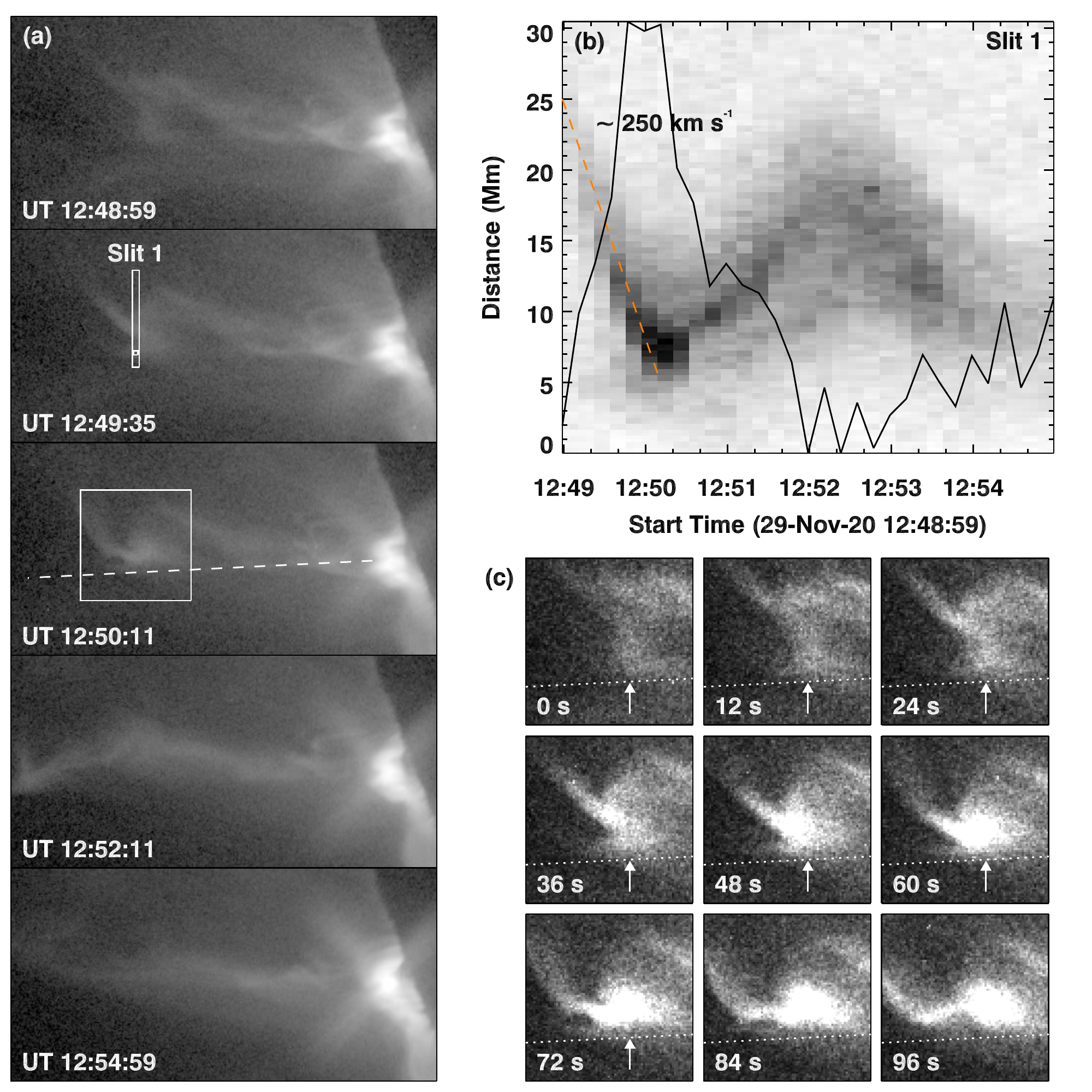}
\caption{Dynamics of the flux rope flank and current sheet seen in the AIA 94\,\AA\ images during the quasi-stable evolution of the current sheet (phase-II; Sect.\,\ref{sec:p2}). Panel (a) shows the AIA 94\,\AA\ time sequence. Slit-1 is the location from where we obtained the space-time map in panel (b). The slanted dashed line in panel (a) is aligned with the current sheet seen in 131\,\AA\ images. The light curve (black) in panel (b) is derived from the small white box embedded in Slit-1. Except for the light curve, panel (b) is same as Fig.\,\ref{fig:p12}e. Panel (c) shows a zoom into the boxed region in panel (a); here 0 s corresponds to UT\,12:48:59. The dotted line is the same as crosses the box in panel (a). See Sect.\,\ref{sec:p2} and Appendix\,\ref{sec:app} for details.\label{fig:p2_94}}
\end{center}
\end{figure*}

In Fig.\,\ref{fig:p12}e, we discussed swaying of flux rope flank and its interaction with the current sheet. In Fig.\,\ref{fig:p2_94} we further demonstrate this interaction between the flux rope flank and the current sheet as seen in AIA 94\,\AA\ images (panel a). As the flank approaches the current sheet, a diffused structure aligned with the 131\,\AA\ current sheet appears also in the 94\,\AA\ images (structure aligned with dotted line in the top row of panel c; pointed by arrows). As the flank moves further southward, the diffused structure becomes narrower and well defined and aligns more clearly with the current sheet (middle row in panel c). A light curve derived directly from this aligned feature shows that it is also brightening as the flank approaches it (panel b). As the flank begins to recede way, intensity of the newly formed thin feature also diminishes (compare the light curve in panel-b and the images at UT\,12:50:11 and UT\,12:52:11 in panel-a). In addition, the flux rope flank sways away from the same plane aligned with the current sheet feature. Here we demonstrated that a section of the current sheet at that location also brightens in the AIA 94\,\AA\ images as the flank approaches it. This close and joint spatial and temporal evolution of the flank and the current sheet indicate that there is indeed likely to be an interaction rather than a superposition of different features along the line of sight. While the 131\,\AA\ feature does not show much change (see Sect.\,\ref{sec:p2}), a section of that feature seen in the 94\,\AA\ images clearly brightens up.

\begin{figure*}
\begin{center}
\includegraphics[width=\textwidth]{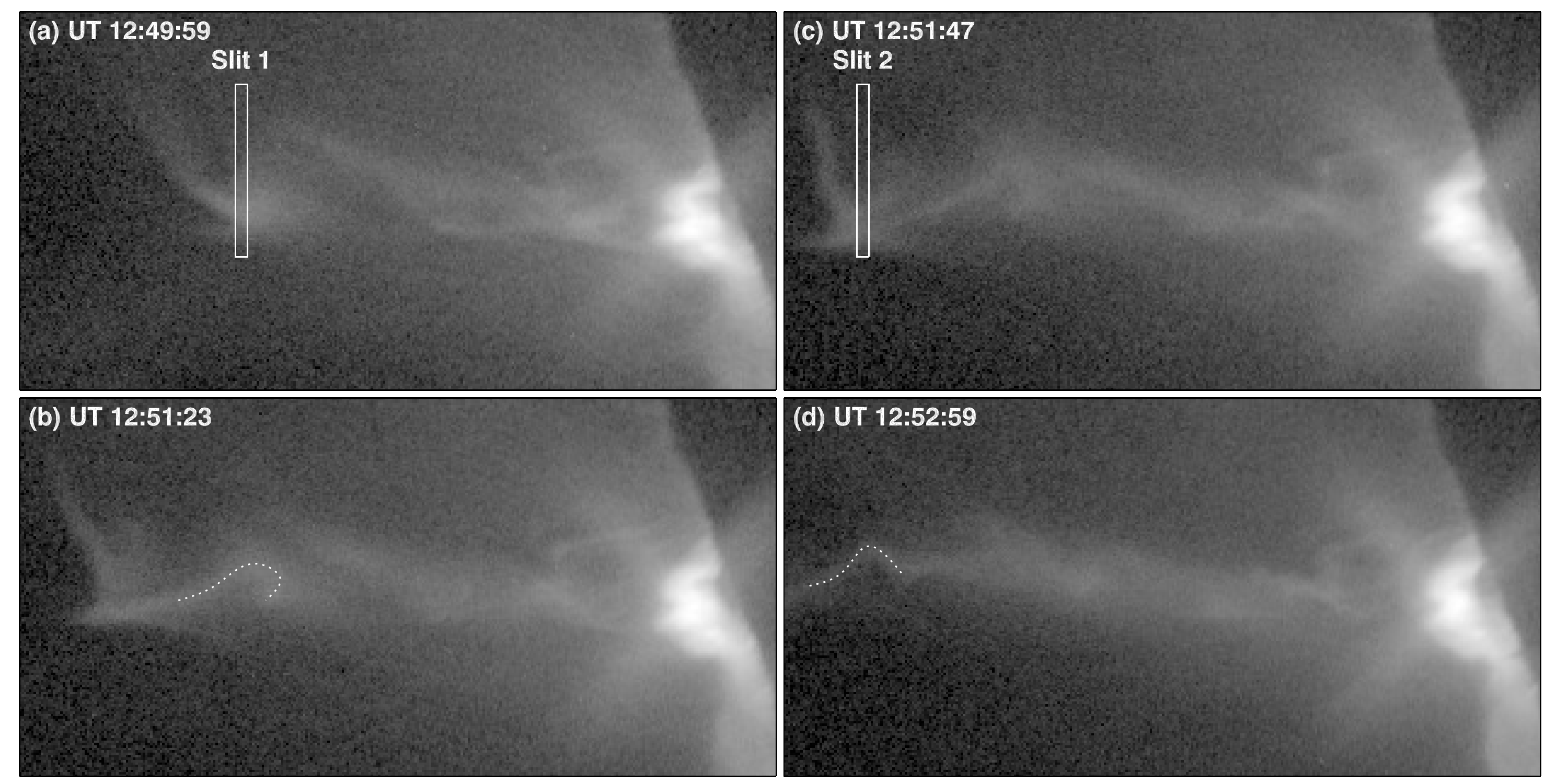}
\caption{Development of kinks in the flux rope flank. Panel (a) shows interaction of the flux rope flank with the current sheet at slit-1. Panel (b) displays a snapshot which highlights the formation of a kink in the flux rope (dotted curve) after its interaction with the current sheet at slit-1. Panels (c) and (d) are the same as panels (a) and (b), but plotted for the interaction episode at slit-2. See also Figs.\,\ref{fig:p12} and \ref{fig:p2_94}. See Sect.\,\ref{sec:p2} and Appendix\,\ref{sec:app} for details.\label{fig:kink_94}}
\end{center}
\end{figure*}

As the flux rope sways away after the interaction with the current sheet at slit-1, it develops a kink feature that section. Similarly, after interaction with the current sheet at slit-2, the flux rope develops another kink feature at that section. Although the flux rope is observed to sway toward the current sheet between slits-1, and -2, it does not develop kink features at these intermediate locations. The kink features are demonstrated in Fig.\,\ref{fig:kink_94}.

\end{document}